\documentclass[twocolumn,showpacs,nofootinbib,10pt,aps,pra,superscriptaddress]{revtex4-1}
\usepackage{amsmath,amssymb,graphicx,color,todonotes,ulem}

\newcommand{\ket}[1]{|#1\rangle}
\newcommand{\bra}[1]{\langle#1|}
\newcommand{\braket}[2]{\langle#1|#2\rangle}

\newcommand{\bc}{\begin{cases}\begin{aligned}}
\newcommand{\ec}{\end{aligned}\end{cases}}
\newcommand{\eq}{\begin{equation}}
\newcommand{\fine}{\end{equation}}
\newcommand{\beq}{\begin{equation}}
\newcommand{\eeq}{\end{equation}}

\newcommand{\uno}{\leavevmode\hbox{\small1\normalsize\kern-.33em1}}
\newcommand{\xx}{{\bf x}_\perp}

\newcommand{\dd}{{\rm d}}

\newcommand{\casi}{\begin{cases}\begin{aligned}}
\newcommand{\casiend}{\end{aligned}\end{cases}}

\newcommand{\re}{\Re {\rm  e}}

\newcommand{\thrms}{\theta_{\rm rms}}
\newcommand{\thEE}{\theta_{\rm EE}}
\newcommand{\dr}{\int^{+\infty}_0\!\!\!\!\!\!\!\!\!\dd r\,\,}
\newcommand{\dphi}{\int^{2\pi}_0\!\!\!\!\dd \phi\,}
\newcommand{\Ftwoone}{\mathcal N}
\renewcommand{\emph}[1]{{\it#1}}
\newcommand{\Mtwo}{\mathcal M^2_{\rm rms}}
\newcommand{\MEE}{\mathcal M^2_{\rm EE}}
\newcommand{\Eqref}[1]{eq. \eqref{#1}}

\begin{document}

\title{A general theorem on the
divergence of vortex beams}

\author{Giuseppe Vallone}\email{vallone@dei.unipd.it}
\affiliation{Department of Information Engineering, University of Padova, via Gradenigo 6/B, 35131 Padova, Italy}
\author{Giuseppe Parisi}
\affiliation{Twist-Off s.r.l., via della Croce Rossa 112, I-35129 Padova, Italy}
\author{Fabio Spinello}
\affiliation{Department of Information Engineering, University of Padova, via Gradenigo 6/B, 35131 Padova, Italy}
\affiliation{Twist-Off s.r.l., via della Croce Rossa 112, I-35129 Padova, Italy}
\author{Elettra Mari}
\author{Fabrizio Tamburini}
\affiliation{Twist-Off s.r.l., via della Croce Rossa 112, I-35129 Padova, Italy}
\author{Paolo Villoresi}
\affiliation{Department of Information Engineering, University of Padova, via Gradenigo 6/B, 35131 Padova, Italy}

\begin{abstract}
The propagation and divergence properties of beams carrying orbital angular momentum (OAM) play a crucial role in many applications.
Here we present a general study on the divergence of optical beams with OAM.
We show that the mean absolute value of the OAM imposes a lower bound on the value of the beam divergence. We 
discuss our results for two different definitions of the divergence, the so called rms or encircled-energy.
 The bound on the rms divergence can be expressed as a generalized uncertainty principle, with applications in long-range communication, microscopy and 2D quantum systems.
\end{abstract}

\maketitle 

\section{Introduction} 
Since the initial work  by Allen {\it et al.}~\cite{alle92pra}, the orbital angular momentum (OAM) of light has attracted increasing interest in multiple fields, including microscopy~\cite{furh05ope}, optical trapping~\cite{grie03nat}, astronomy~\cite{anzo08AA,mari12ope},  radio~\cite{tamb12njp} and optical communication~\cite{wang12npho,bozi13sci,vall14prl,mirh15njp}
and fundamental physics~\cite{mera09nph,tamb11nph,damb14prl}.
OAM beams are characterized by a singular phase factor $\exp(i \ell \phi)$, where $\phi$ refers to the azimuthal angle around the beam axis and the topological charge, $\ell$, is an integer parameter representing different OAM values~\cite{alle92pra}. 
When $\ell\neq0$, the beam 
presents an optical vortex on its axis due to
the phase singularity.

The study of the propagation and divergence properties of OAM beams play a crucial role in many applications, and in particular in long-range communication systems~\cite{yan14nco,tamb15ras}
and microscopy. 
In this context, standard Laguerre-Gaussian (LG) beams have already been studied in 
detail~\cite{phil83apo,sieg91ieee,pari14ope, padg15njp, redd15aop}. 
Recently~\cite{vall15opl}, a preliminary analysis
was also carried out on the Circular Beam (CiB), which represents a general analytical solution of the
paraxial wave equation with OAM~\cite{band08opl,vall15opl}. 
Indeed, several well known beams carrying OAM
-- such as the standard~\cite{sieg86lasers} or elegant~\cite{wuns89josa} LG
beams, Bessel-Gauss 
beams~\cite{guti05josa}, optical vortex beams~\cite{berr04joa} and others~\cite{band08opl} --
are particular cases of CiBs obtained by setting  specific values to the beams' parameters.
Moreover, CiBs naturally arise when $q$-plates~\cite{marr06prl}
or phase plates~\cite{beij94opc} are used to general OAM from a Gaussian beam~\cite{kari09ope,vall15opl}.

The divergence properties of generic incoherent superposition of LG modes can be easily bounded
by knowing the divergence angles of the Laguerre-Gauss beams. 
However, for generic (coherent) beams -- that can be always expressed as a coherent superposition of LG modes -- there are no known bounds.
One could image that by coherently adding different LG modes 
with the same OAM it could be possible to lower the divergence of the beam, by keeping fixed the value of the OAM.
A central question is thus the following: by fixing the OAM content and
arbitrarily changing the radial profile of a beam
is it possible to reduce its divergence?
The above question is crucial for applications that require the optimization of the far-field propagation or the focusing properties of the used beam.

Here we answer to the above question by presenting a general study on the propagation and divergence properties of OAM beams, 
including CiBs as special case. 
The study on the divergence of optical beams will led to a  formulation of an uncertainty principle,
based on the mean value of the OAM.
Our main result can be stated as follows: for any optical monochromatic paraxial beam with a mean value of OAM given by $\langle |\ell|\rangle$, the product of the spatial extent, $\sigma_r$, and the spread $\sigma_k$ in the wavector-space
 is lower bounded by 
\beq
\label{fundamental}
\sigma_k\sigma_r\geq 1+\langle |\ell|\rangle\,.
\eeq
This result can be heuristically explained by noticing that, for an optical beam with $\ell$ units of OAM, 
the Poynting vector angle with respect to the 
propagation axis is given by $\ell/(k r)$~\cite{alle20opc,leac06ope}. Then, the ``spreading'' of the Poynting vectors, related to
the beam divergence, tends to increase with
$\ell$. 
Our result is a kind of {\it no-go theorem}: any optimization of the beam radial profile cannot improve the
divergence below the limit imposed by \Eqref{fundamental}.

We now derive \Eqref{fundamental}
and show how the inequality 
could be also exploited in microscopy and
formulated as a position-momentum uncertainty principle.

\begin{figure}[t]
\includegraphics[width=7cm]{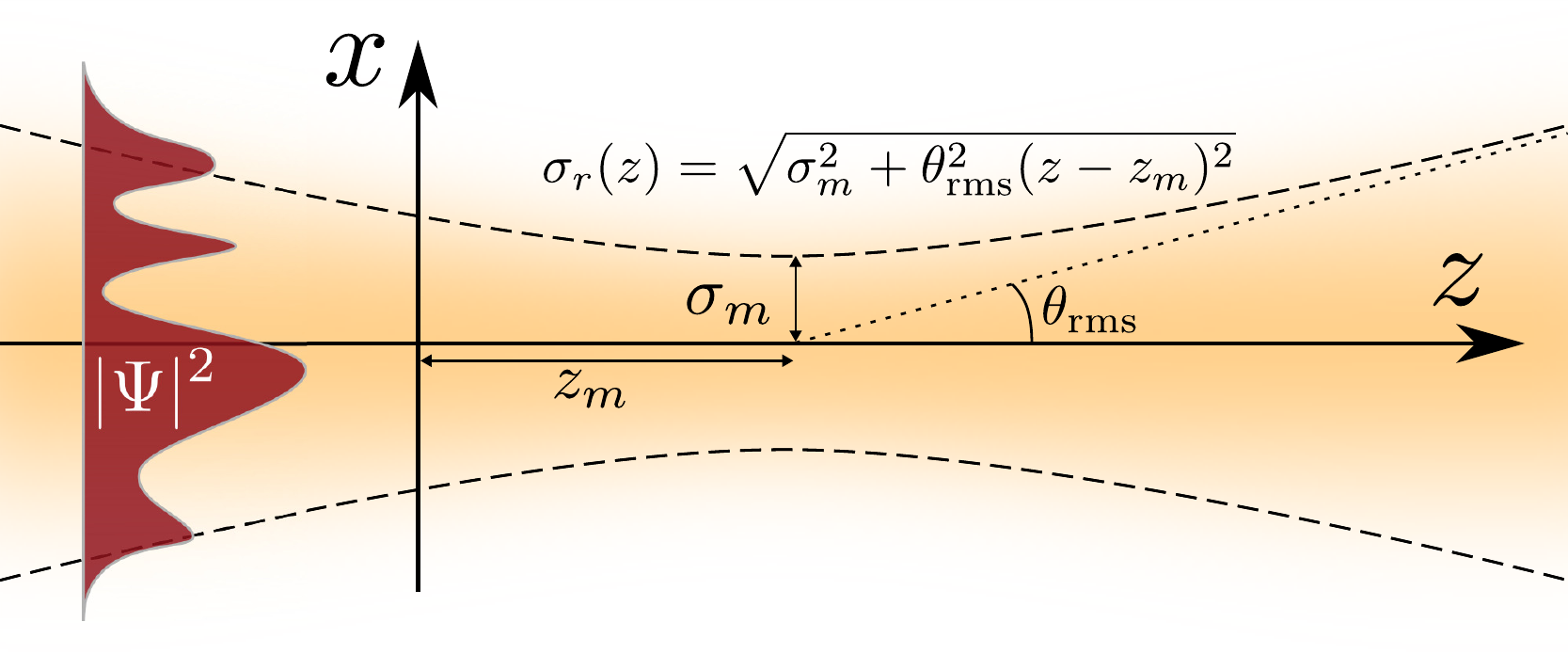}
\caption{(Color online) Pictorial representation
of the rms divergence $\theta_{\rm rms}$ of a generic
paraxial beam $\Psi(x,y,z)$ 
propagating along the $z$ axis.
Dashed lines represent $\sigma_r(z)$.
The parameters $\sigma_m$ and $z_m$ are the minimum
value of $\sigma_r(z)$ and its location
respectively.}
\label{fig1}
\end{figure}

\section{Bounding the rms divergence}
Let's consider a generic monochromatic and
paraxial beam propagating in the $z$ direction: 
its wavefuction can be written as $\Psi(x,y,z)e^{i\omega t}$,
where $\omega$ is the angular frequency and $k=\omega/c$ the
wavenumber.
By defining $\xx=(x,y)=(r,\phi)$ as the coordinate
on the plane transverse to propagation, without loss of generality we can consider only beams
with $\langle \xx(z)\rangle=0$~\cite{foot1}.
The starting point for our analysis is the so called root-mean-square (rms) {\it  far field angle of divergence} $\theta_{\rm rms}$. 
This is defined, for a generic paraxial beam $\Psi$, as 
\beq
\label{thrms}
\theta_{\rm rms}=\lim_{z\rightarrow\infty}\frac{\sigma_r(z)}{z}
\,,
\eeq
where $\sigma^2_r(z)$ is the variance of the intensity
$I(r,\phi,z)=|\Psi(r,\phi,z)|^2$ given by:
\beq
\label{srms}
\sigma^2_r(z)=\int\dd \phi\int\dd r\, r^3 
I(r,\phi,z)\,.
\eeq
The value of $\sigma_r(z)$ represents the beam radius at location $z$~\cite{phil83apo,cart80apo}~\cite{foot2}.
As underlined by the r.h.s. of \Eqref{thrms}, the far field angle of divergence $\theta_{\rm rms}$ quantifies the asymptotic rate of variation of  $\sigma_r(z)$.

The divergence $\thrms$ can be explicitly evaluated
by exploiting the expansion 
of $\Psi$ in LG modes, 
a complete (orthonormal) basis set of solutions of the paraxial
wave equation. 
Due to the completeness of the LG basis, any paraxial beam $\Psi$ can be expanded as:
\beq
\label{LGexp}
\Psi(r,\phi,z)=\sum_{n=0}^{+\infty}
\sum_{\ell=-\infty}^{+\infty}\psi_{n,\ell}\, 
{\rm LG}_{n,\ell}(r,\phi,z)\,.
\eeq
In the above expression the (complex) expansion  coefficients $\psi_{n,\ell}$ are adimensional 
and they 
are normalized such that
$\sum_{n,\ell}|\psi_{n,\ell}|^2=1$.
The waist parameter $w_0$ of the LG modes
determines the physical size of $\Psi$ (see appendix A).
The integers $n$ and $\ell$ respectively represent the 
radial quantum number and the OAM content of each LG mode.

{
As first derived by Siegman in~\cite{sieg91ieee},
the square of the beam radius $\sigma^2_r(z)$ of a generic beam has a parabolic dependence on $z$, namely
$\sigma^2_r(z)=\sigma^2_m+\theta^2_{\rm rms}(z-z_m)^2$. 
The  parameters
$\sigma_m$ and $z_m$ represent the minimum
value of $\sigma_r(z)$ and its location on the $z$-axis respectively.  A pictorial representation of
$\sigma_r(z)$ is given in Fig. \ref{fig1}.
As detailed in Appendix \ref{appendixA},
by exploiting the LG expansion of eq. \eqref{LGexp}, the parameters $\sigma_m$, $\theta_{\rm rms}$ and $z_m$
can be related to the coefficients $\psi_{n,\ell}$.
We note that
$\sigma_m$ may be considered as a free parameter
that determines the physical transverse size of the beam. Indeed, since $w_0$ and
$\sigma_m$ are related through a combination of
the coefficients $\psi_{n,\ell}$,
by suitably tuning $w_0$ it is possible 
to arbitrarily choose $\sigma_m$.
}


By multiplying $\sigma_m$, $\thrms$ and $k$ it is possible to obtain an adimensional quantity, the
so called {\it beam quality} $\Mtwo$-factor~\cite{Saleh-Teich},
which is independent on the physical
size of the beam.
{ By exploiting the LG expansion of eq.  \eqref{LGexp},
the $\Mtwo$-factor of a generic beam can be expressed as:}
\beq
\label{M2}
\begin{aligned}
\Mtwo\equiv k\theta_{\rm rms}\sigma_m
=\sqrt{\alpha^2-|\beta|^2}\,,
\end{aligned}
\eeq
where $\alpha$ and $\beta$ depend only on the
expansion coefficients $\psi_{n,\ell}$
of \Eqref{LGexp}. Their
explicit expressions are the following:
\beq
\begin{aligned}
\alpha&=1+\langle |\ell|\rangle+\Phi\,,
\\
\beta&=\sum_{n,\ell}2\sqrt{n(|\ell|+n)}\psi_{n,\ell}\psi^*_{n-1,\ell}\,,
\end{aligned}
\eeq
with 
$\langle|\ell|\rangle=\sum_{n,\ell}|\ell||\psi_{n,\ell}|^2$ and
$\Phi=\sum_{n,\ell}2n|\psi_{n,\ell}|^2$.
As suggested by the notation, $\langle|\ell|\rangle$
represents the mean absolute value of the 
OAM of the beam. 

The value of $\langle|\ell|\rangle$ can be
used to bound
the rms divergence.
As demonstrated in appendix \ref{appendixB}, the
factor $\sqrt{\alpha^2-|\beta|^2}$ is lower bounded
by $1+\langle|\ell|\rangle$. 
Then, the main result
of our analysis can be summarized by the following bound:
\beq
\label{bound}
\begin{aligned}
{ \mathcal M^2_{\rm rms}\geq 1+\langle|\ell|\rangle\,,}
\end{aligned}
\eeq
implying that the mean absolute value
of the OAM increases the value of the beam divergence.
{ We note that 
 the well known~\cite{debr67ine,hogson-Weber}  inequality $\mathcal M^2_{\rm rms}\geq 1$
for generic beams has no contribution of the orbital angular momentum as in eq. \eqref{fundamental}.
Moreover, while it was well known that for a LG$_{n,\ell}$ mode the beam quality factor is given by
$\mathcal M^2=2n+|\ell|+1$, it was not known what happens for a generic coherent superposition.
We would like to point out that
the bound was only known for {\it incoherent} superposition of LG modes~\cite{sieg90spie}
(with incoherent superposition we denote a beam whose intensity is given by
$I(x)=\sum_{n,\ell}|\psi_{n,\ell}|^2 |{\rm LG}_{n,\ell}(x)|^2$).
Only in this case, the $\mathcal M^2$ factor is trivially bounded by 
$1+\langle|\ell|\rangle$ since for incoherent superposition we have
$\mathcal M^2=\sum_\ell\sum_n (2n+|\ell|+1)|\psi_{n,\ell}|^2\geq 1+\langle|\ell|\rangle$.
One could ask the question
whether by {\it coherently} adding different LG modes it is possible to reduce the divergence of the beam up 
to the standard bound $\mathcal M^2_{\rm rms}\geq 1$.}
As shown by eq. \eqref{bound}, the answer is no, since the mean value of the OAM increases the lower bound on $\mathcal M^2$.

The bound in eq. \eqref{bound} is tight, since it can be
achieved by arbitrary superposition of LG modes with $n=0$.
Indeed, for generic superposition
$\sum_\ell \varphi_\ell {\rm LG}_{0,\ell}$ we have
$\Mtwo=1+\sum_\ell|\varphi_\ell|^2|\ell|=1+\langle|\ell|\rangle$,
saturating the bound given by \Eqref{bound}.

This result can be easily converted to an 
Heisenberg-like uncertainty principle (see \Eqref{fundamental}).
Indeed, the rms divergence $\theta_{\rm rms}$ is related
to the standard deviation of the wavevectors.
By defining the Fourier transform of the field as
$\widetilde \Psi(k_x,k_y)=\frac{1}{2\pi}\int\dd x\dd ye^{i(k_xx+k_yy)}\Psi(x,y,0)$, 
it is well known that, at large $z$, 
$\Psi(x,y,z)\sim \frac{ik}{z}
e^{ik(z+\frac{r^2}{2z})}
\widetilde\Psi(\frac{k}{z}x,\frac{k}{z}y)$~\cite{Saleh-Teich}.
Then, at large $z$, the radial variance can be
approximated to
$\sigma^2_r(z)\sim\frac{z^2}{k^2}\sigma^2_{k}$
with $\sigma^2_{k}=\int \dd k_x \dd k_y (k_x^2+k_y^2)
|\widetilde \Psi(k_x,k_y)|^2$ the variance in the Fourier space.
By using the definition \eqref{thrms}
and by noticing that $\sigma_r\geq\sigma_m$ by definition, \Eqref{bound} can be then rewritten as
\beq
\label{UPk}
\sigma_r\sigma_k\geq 1+\langle|\ell|\rangle\,.
\eeq

As already said, the bound can be applied also
to microscopy. Indeed, we may exploit the
well-known relationship between the Fourier transform
and the images in the focal plane of thin lenses~\cite{Saleh-Teich}. 
In the focal plane the intensity is given by
$|\frac{k}{f}\widetilde\Psi(\frac{k}{f}x,\frac{k}{f}y)|^2$.
Then, the spatial extent in the focus is given
by $\sigma_{\rm focus}=\frac{f}{k}\sigma_k$ such that
\beq
\sigma_{\rm source}\sigma_{\rm focus}\geq \frac{f\lambda}{2\pi}(1+\langle|\ell|\rangle)\,,
\label{micro}
\eeq
limiting the dimension of the focused spot when
the beam carries OAM. We note that
the above relation should be taken into account
in the coupling of OAM beam into optical fibers.

Finally, the bound \eqref{bound} corresponds to an improved
Heisenberg uncertainty principle.
Indeed, the paraxial wave equation is mathematically
equivalent to the 2D-Schr\"odinger equation for a free particle:
the direction of propagation $z$ becomes the
time evolution parameter $t$ for the free particle
and the wavevector $k$ is related to the particle
momentum by $k=p/\hbar$. 
Equation \eqref{UPk} is then equivalent to 
\beq
\sigma_r\sigma_p\geq \hbar(1+\langle|\ell|\rangle)\,,
\label{scho}
\eeq
for a free-particle in 2D carrying OAM.

Eqs. \eqref{UPk}, \eqref{micro} and \eqref{scho} are equivalent relations that show that
the OAM becomes a fundamental quantity to study different problems,
from the properties of optical beams
in long-distance propagation, microscopy and optical
fiber coupling
to the behavior of 2D quantum free-particles.

\begin{figure}[t]
\includegraphics[width=8cm]{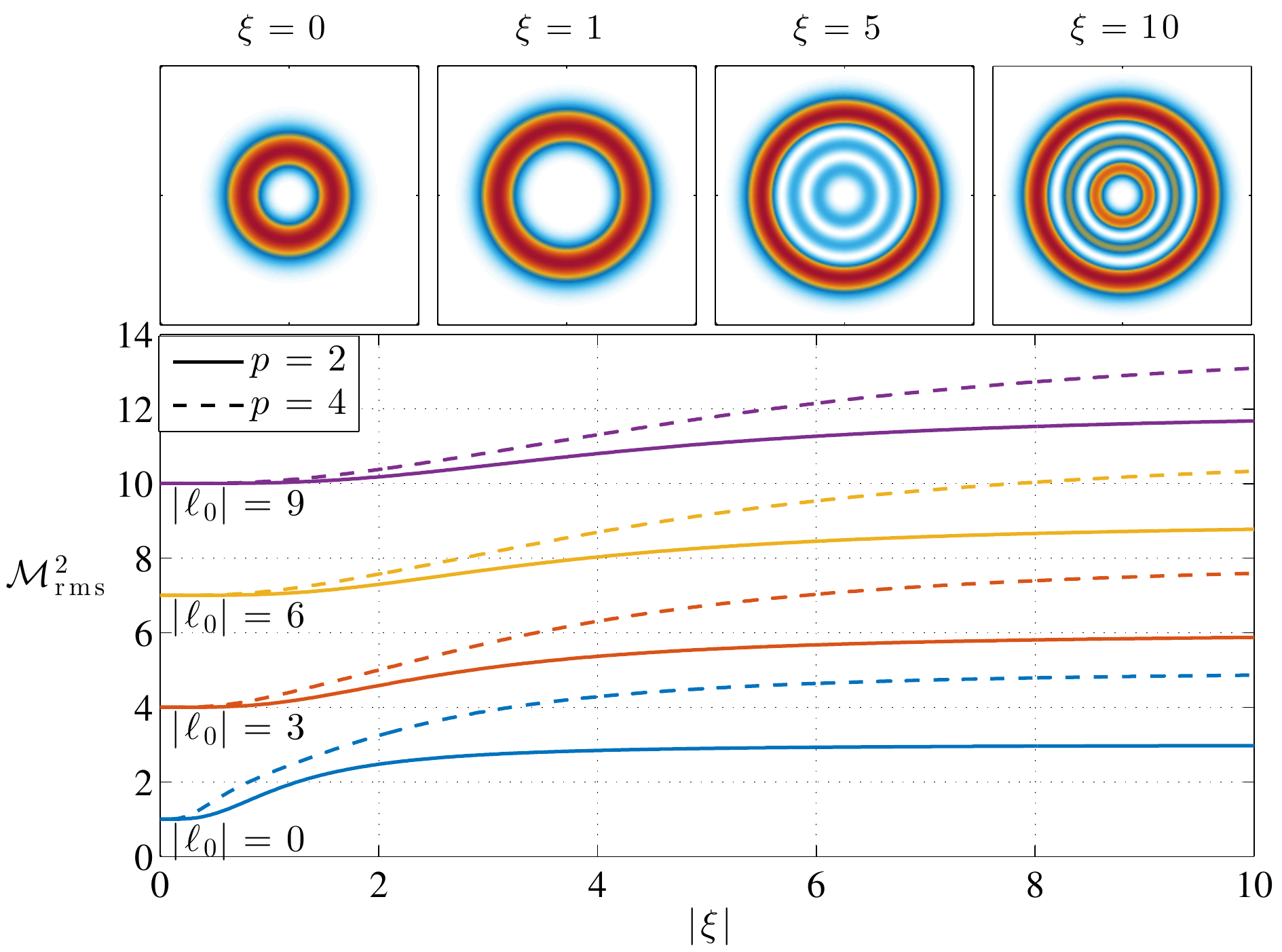}
\caption{(Color online) $\Mtwo$-factor 
 for the CiBs with different value of $|\xi|$, $|\ell_0|$ and $p$. The lower bound 1+$|\ell_0|$
 is obtained when $\xi=0$: in this case 
 the CiB reduces to a
 LG$_{0,\ell_0}$ mode. { In the inset we show the intensity patterns of different CiBs with $\ell_0=3$ and $p=4$.}}
\label{thrms_CiBs}
\end{figure}

\subsection{CiB case} As an example, we now explicitly evaluate the $\mathcal M^2_{\rm rms}$ factor of  the Circular Beams. 
A generic CiB is determined by three complex parameters $\xi$, $q_0$ and $p$ and one 
integer parameter $\ell_0\in\mathbb Z$.
The parameter  $\xi$ is related to the beam ``shape'' as illustrated in Fig. \ref{thrms_CiBs}.
Specific values of $\xi$ identify some well-known beams:
for instance, the limit $\xi\rightarrow+\infty$ corresponds to the LG modes,
while CiBs with $|\xi|=1$ correspond to the generalized Hypergeometric-Gaussian modes~\cite{kari07opl,kari09opl,vall15opl}.
The parameter $q_0$ is related to the physical scale (similarly to the complex beam parameter of the Gaussian beam~\cite{sieg86lasers}).
Finally, $p$ is a radial index and $\ell_0$ corresponds to the carried OAM.
For a circular beam,
the expansion in term of LG mode 
 was derived in~\cite{vall15opl} and it is written as
\beq
{\rm CiB}^{(q_0,\xi)}_{p,\ell_0}(r,\phi,z)=
\sum_{n=0}^{+\infty}\sum_{\ell=-\infty}^{+\infty}
\psi_{n,\ell}\, 
 {\rm LG}_{n,\ell}(r,\phi,z)\,,
\eeq
with
\beq
\label{psiCiB}
\psi_{n,\ell}=\delta_{\ell,\ell_0}\frac{\xi^n}{
\sqrt{\Ftwoone}}
\frac{\Gamma(n-\frac p2)}{\Gamma (-\frac p2)}
\sqrt{\frac{|\ell_0|!}{n!(|\ell_0|+n)!}}\,.
\eeq
In the previous equation
 $\Ftwoone$ is a normalization factor given by
the Hypergeometric function $\Ftwoone=\!\ _2F_1[-\frac p2, -\frac{p^*}{2}, 1 + |\ell_0|,|\xi|^2]$.
For simplicity, in $\psi_{n,\ell}$ we did not explicitly indicate the dependence on $p$, $\ell_0$ and $\xi$.

The parameters $\Phi$ and $\beta$ that allow to calculate the beam quality factor can be explicitly
evaluated from \eqref{psiCiB} and their value is given by
$\Phi
=\frac{1}{\Ftwoone}\frac{|p\, \xi|^2}{2+2|\ell_0|}\ _2F_1[1-\frac p2, 1-\frac{p^*}{2}, 2 + |\ell_0|,|\xi|^2]$
and 
$\beta=\xi(\Phi-p)$.
The mean absolute value of the OAM is simply $\langle|\ell|\rangle=|\ell_0|$. Then, the $\Mtwo$-factor
depends only on $|\xi|$, $p$ and $|\ell_0|$
as
\beq
\Mtwo=\sqrt{(1+|\ell_0|+\Phi)^2-|\xi(\Phi-p)|^2}.
\eeq
The behavior of $\Mtwo$ is shown in Fig. \ref{thrms_CiBs} 
for different values of $\ell_0$ and $p=2$ or $p=4$.
In particular, when $|\xi|=1$, the $\Mtwo$-factor has a simple expression:
\beq
\label{M2simple}
\Mtwo=\sqrt{(1+|\ell_0|)^2+\frac{|p|^2}{\re(p)+|\ell_0|}}\,.
\eeq
Since the rms divergence can be defined in this case only when $\re(p)>-|\ell_0|$ (see~\cite{vall15opl}),
it can be easily checked that quality factor expressed in eq. \eqref{M2simple} satisfies the general bound of \Eqref{bound}.

\section{Encircled-Energy divergence}
As already noted in~\cite{vall15opl}, 
the CiBs with $|\xi|=1$ and 
$-|\ell_0|-1<\re(p)\leq-|\ell_0|$ are square 
integrable beams but their 
rms divergence cannot be defined.
This feature is
common to all fields whose intensity
(at large $r$ and fixed $z$) fall-off as $1/r^{2+a}$, with $0<a\leq2$. For this reason, 
in these cases an alternative expression of the divergence should be used.
To this purpose, we may define the so-called 
{\it Encircled-Energy far field angle of divergence} $\theta_{\rm EE}$
by
\beq
\theta_{\rm EE}=\lim_{z\rightarrow+\infty}\frac{R_{\rm EE}(z)}{z}\,,
\eeq
where $R_{\rm EE}(z)$ is the 
 {\it Encircled-Energy radius}.
 $R_{\rm EE}(z)$ must be calculated by 
the implicit relation $\int\dd\phi \int^{R_{\rm EE}(z)}_0\dd r\, r I(r,\phi,z)=E_0$, with 
$E_0$ a fixed constant.
Here, 
$R_{\rm EE}(z)$ corresponds
to the radius 
whose corresponding circle centered on the beam axis contains a given fraction $E_0$ of the total beam energy. By definition, the
divergence $\thEE$ is well defined for any square integrable beam.

\begin{figure}[t]
\includegraphics[width=8cm]{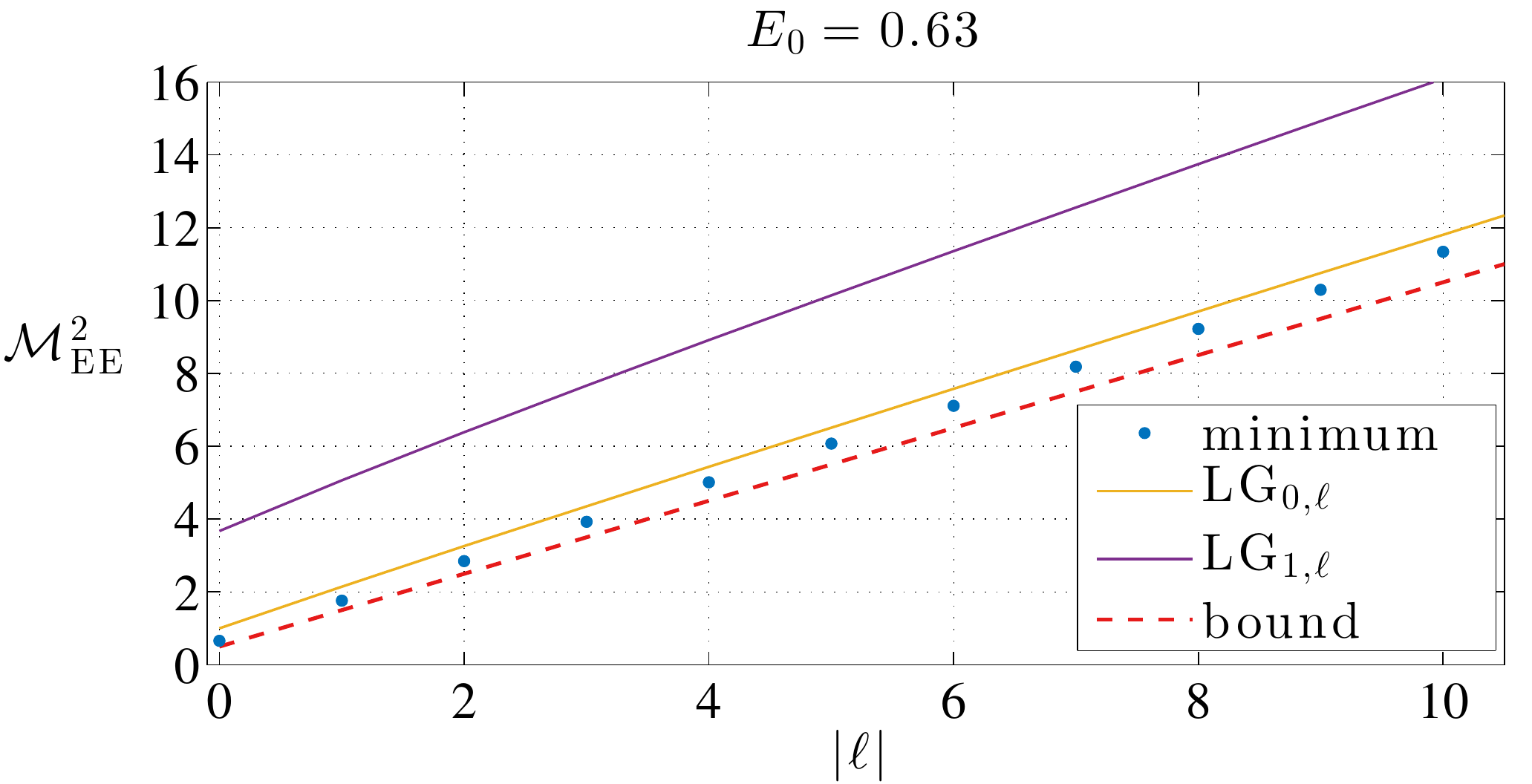}
\caption{(Color online) Minimum values of $\MEE\equiv k\thEE R_m$ calculated
over random beams with fixed value of $\ell$.
We used $E_0\simeq 0.63$. 
We compare the minimum
values of $\MEE$ with the
ones obtained with the ${\rm LG}_{0,\ell}$ and
${\rm LG}_{1,\ell}$ modes.
Dashed line represents the lower bound
as given in \Eqref{bound_thEE}.}
\label{kthR}
\end{figure}
We now show that a bound similar to
\eqref{bound} can be obtained for $\thEE$,
with $\sigma_m$ replaced by $R_m\equiv \min_z R(z)$,
the minimum of the encircled-energy radius.
As it happens for the
rms divergence, the product 
$\MEE\equiv k\thEE R_m$ is adimensional and
depends only on the expansion coefficients $\psi_{n,\ell}$. As detailed in appendix \ref{appendixB}, it is possible
to show that
\beq
\label{MEE}
\MEE=\sqrt{{T_\infty}{\min_Z[(1+Z^2)T(Z)]}}\,,
\eeq
with  $T_\infty\equiv\lim_{Z\rightarrow \infty}T(Z)$.
The function $T(Z)$ is defined only in
terms of the expansion coefficient by
the implicit relation
\beq
E_0=\sum_\ell\int_0^{T(Z)}
|\mathcal U_\ell(t,Z)|^2\dd t\,,
\eeq 
where
$\mathcal U_\ell(t,Z)=
\sum_{n=0}^\infty
\sqrt{\frac{ n!\, t^{|\ell|}}{(|\ell|+n)!}} \psi_{n,\ell}
{(\frac{1+iZ}{1-iZ})}^n e^{-\frac{t}{2}}
L_n^{(|\ell|)}(t)$.

We numerically evaluated the
minimum of 
$\MEE$ 
for a fixed value of OAM $\ell$. 
For each value of $\ell$,
we searched for the minimum of $\MEE$ 
by using a truncated superposition (up to $n=10$) of the ${\rm LG}_{n,\ell}$ modes as in \Eqref{LGexp}.
We fixed $E_0=1-1/e\simeq0.63$:
such value is required to achieve $\thEE=\thrms$ for the Gaussian beams (i.e.
the ${\rm LG}_{0,0}$ mode). Different values of $E_0$ will be discussed later.
The results of the
numerical minimization, performed by
using the Nelder-Mead algorithm~\cite{neld65tcj},
are shown in Fig. \ref{kthR} for
different values of $\ell$. 
For comparison, we also show the value of $k\thEE R_m$ for the 
${\rm LG}_{0,\ell}$ and ${\rm LG}_{1,\ell}$ modes.
The minimum of $k\thEE R_m$ for a random beam is
slightly lower than the value obtained for the
${\rm LG}_{0,\ell}$ mode, but it
cannot be arbitrarily low. 
Indeed, our numerical minimization shows 
that the 
divergence $\thEE$ satisfies the following
bound for any beam with a fixed value of
OAM given by $\ell$:
\beq 
\label{bound_thEE}
\thEE\geq \frac{1}{k R_m}(c_0+|\ell|)\,,
\eeq 
with $c_0=0.5$.
In figure Fig. \ref{kthR} we report such  bound  by using a dashed line.

\begin{figure}[t]
\includegraphics[width=8cm]{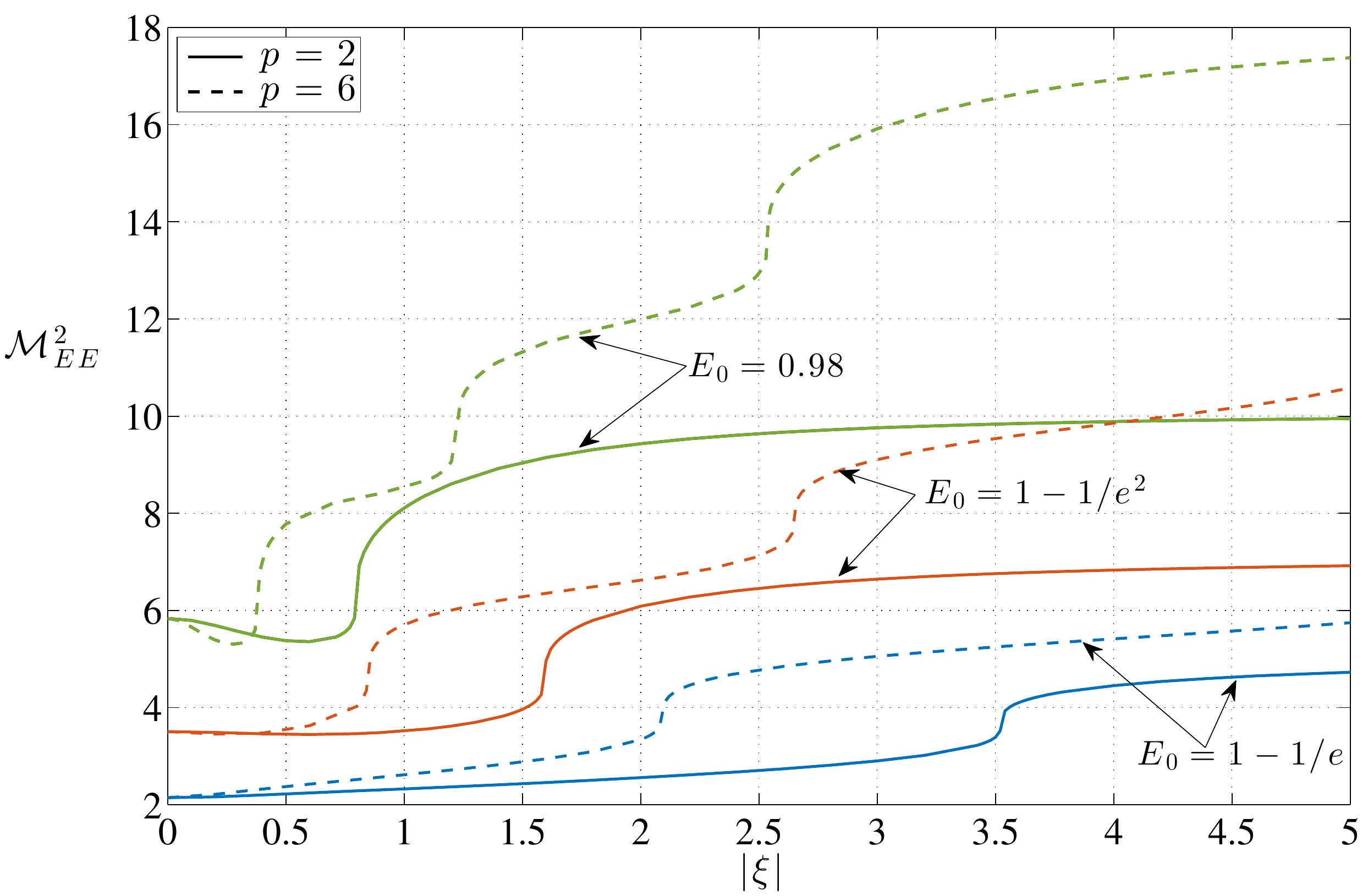}
\caption{(Color online) Encircled-Energy divergence of CiBs for
different values of $E_0$. We plot
the product $\mathcal M^2_{\rm EE}\equiv k\thEE R_m$ in function of $|\xi|$ (for real $\xi$) with $\ell=1$ and $p=2$ or $p=6$.
By increasing $E_0$, the value of $\mathcal M^2_{\rm EE}$ increases.
}
\label{cibsEE}
\end{figure}

The value $E_0\simeq0.63$ may seem arbitrary and indeed the value of the product $k\thEE R_m$ depends on such choice. For instance, as
illustrated in Fig. \ref{cibsEE} for the CiBs,
by increasing the value of $E_0$, the
value of $\mathcal M^2_{\rm EE}$ will increase.
In order to check the validity of the bound in \Eqref{bound_thEE}, we tested it for different values of $E_0$,
obtaining the same bound \eqref{bound_thEE} with
different values of $c_0$: in particular,
for $E_0=1-1/e^2\simeq0.86$ (a common value used to define the divergence) we obtained $c_0=1.8$ while for 
$E_0=0.98$ we obtained $c_0=3$. 
Details of the minimization procedure
and the results for $E_0=0.86$ and $E_0=0.98$ are
presented in appendix \ref{appendixC}. {In appendix \ref{appendixD} we also show some
examples of beams that minimize $\mathcal M^2_{\rm EE}$.}

We conjecture that, as long as $E_0>0.5$, 
the bound \eqref{bound_thEE} holds.
The parameter $c_0$ will depends on 
the specific value chosen for $E_0$.
The conjecture is based on the fact that, at large $|\ell|$, the asymptotic divergence of the
LG$_{0,\ell}$ modes is given by
$\mathcal M^2_{\rm EE}\sim|\ell|+t\sqrt{2|\ell|} $
with $t={\rm erf}^{-1}(2E_0-1)$ (see details in appendix \ref{appendixC}). When $E_0>0.5$ we have 
$t>0$ and the $\mathcal M^2$ parameter for the LG$_{0,\ell}$
is always larger than $|\ell|$.

\section{Conclusions} 
We studied the divergence of
generic beams carrying OAM. We demonstrated that the
rms divergence is bounded by the absolute
mean value of the beam orbital angular momentum
(see \Eqref{bound}). 
We have shown that such bound provides  an uncertainty relation that is useful for applications involving long-distance propagation of beams and in the study of vorticity generation, in the limits of imaging and illumination in microscopy as well as in
the study of quantum free-particle confined in 2D
(see \Eqref{UPk}, \eqref{micro} and \eqref{scho}).
Finally, we demonstrated that a similar bound
holds for a different definition of the
divergence, the so called Encircled-Energy divergence.
Our results prove that the presence of orbital angular momentum enhances the angular spreading of the
beam, leading to an increase of the uncertainty 
relation between the spatial and wavevector extent. Such uncertainty 
implies an increased beam divergence
and a lower ability in focusing the beam.

{ Our results are obtained for paraxial beam. A recent work~\cite{alon11jop} showed that a similar bound,
$kR_{\rm rms} \sin\theta_{\rm rms} \geq \cos\theta_{\rm rms}+|\langle L_z\rangle|$,
can be achieved for non-paraxial electromagnetic beams. 
However, the bound was shown only for a combination of two eigenstates of 
the OAM operator $L_z=-i\frac{\partial}{\partial \phi}$. We note that the result obtained in~\cite{alon11jop}
reduce to eq. \eqref{bound} for OAM eigenstates and small $\theta$ (i.e. paraxial beam).
Our technique, combined to the results obtained in~\cite{alon11jop} 
may allow to investigate the divergence properties of generic non-paraxial beams carrying OAM.
We leave such investigation for future works.}

\acknowledgements
G.V. and P.V. acknowledge the Strategic-Research-Project QUINTET of the Department of Information Engineering, University of Padova, and the Progetto di Ateneo PRAT 2013 ``OAM in free space: a new resource for QKD'' (CPDA138592).
G.P., E.M., F.S. and F.T. acknowledge the support and collaboration
of SIAE Microelectronics.

\appendix

\section{Evaluation of $\sigma_r(z)$}
\label{appendixA}
In this section we analytically evaluate  the 
beam variance $\sigma^2_r(z)$ and the rms divergence 
$\theta_{\rm rms}$ of a generic beam.
We consider a generic paraxial beam $\Psi(r,\phi,z)$ propagating in the $z$ direction.
The beam can be expanded in the basis of the LG modes as follows:
\beq
\label{LGexp_SI}\Psi(r,\phi,z)=
\sum_{n=0}^{+\infty}
\sum_{\ell=-\infty}^{+\infty}
\psi_{n,\ell}\, 
{\rm LG}_{n,\ell}(r,\phi,z)\,,
\eeq
with $\sum_{n=0}^{+\infty}
\sum_{\ell=-\infty}^{+\infty}
|\psi_{n,\ell}|^2=1$. 

Our convention for the (normalized)
LG modes with beam waist parameter $w_0$ is the following:
\beq 
\begin{aligned}
{\rm LG}_{n,\ell}=&\sqrt{\frac{2}{\pi}}
\sqrt{\frac{ n!}{(|\ell|+n)!}}
\frac{e^{-\frac{ikr^2}{2q(z)}}}{w(z)}{\left(\frac{\sqrt{2}r}{w(z)}\right)}^{|\ell|}
\times\\
&L_n^{(|\ell|)}\!\!\left(\frac{2 r^2}{w^2(z)}\right)e^{i\ell\phi}e^{i(2n+|\ell|+1)\zeta(z)}\,,
\end{aligned}
\eeq
where $n$, $\ell\in \mathbb Z$ with $n\geq 0$, $L_n^{(|\ell|)}(x)$  is the generalized Laguerre polynomial,
$w(z)=w_0\sqrt{1+(z/z_0)^2}$ is the beam size, $z_0=kw_0^2/2$ is the Rayleigh range and
$\exp[i\zeta(z)]=(z_0+i z)/|z_0+i z|$ is the Gouy phase. 
The integers $n$ and $\ell$ respectively represent the 
radial quantum number and the OAM content of each LG mode.
We note that the physical scale of the beam $\Psi$ in \Eqref{LGexp_SI} is determined by the value of the
beam waist parameter $w_0$.

The rms variance 
$\sigma_r^2(z)$ of the beam $\Psi$ is 
defined in eq. (3) of the main text. 
By using the LG mode expansion, the variance may 
be rewritten as:
\begin{align}
\label{srms_SI}
\sigma^2_{r}(z)&=
\dphi\dr
r^3 |\Psi(r,\phi,z)|^2
\\
\notag
&=\sum_{n,n',\ell,\ell'}\psi_{n,\ell}
\psi^*_{n',\ell'}\!\!
\dphi\!\!\dr
r^3{\rm LG}_{n,\ell}
{\rm LG}^*_{n',\ell'}\,.
\end{align}
The above integral on the LG modes is evaluated as~\cite{vall15opl}:

\begin{align}
&\dphi\dr
r ^3 \,\text{LG}_{n,\ell}(r,\phi,z)  \text{LG}^*_{n',\ell'}(r,\phi,z)=
\\
\notag
&\frac{\delta_{\ell,\ell'}}2 [
B_{\ell,n}(z) \delta _{n',n}
- C_{\ell,n}(z) \delta _{n'+1,n} -C^*_{\ell,n'}(z) \delta _{n',n+1} 
]\,,
\end{align}
with $B_{\ell,n}=w^2(z)(|\ell|+2n+1)$ and 
$C_{\ell,n}=w^2(z)e^{2 i \zeta(z)}\sqrt{n (|\ell|+n)}$. 

By plugging the above result into eq. \eqref{srms_SI}
it is possible to obtain an explicit expression for $\sigma^2_{\rm rms}(z)$, namely:
\beq
\begin{aligned}
\sigma^2_{\rm rms}(z)
&=\frac{w^2(z)}2\left [1+\langle|\ell|\rangle+\Phi-\re(e^{2 i \zeta(z)}\beta)\right]\,,
\end{aligned}
\eeq
where $\langle|\ell|\rangle$, $\Phi$ and $\beta$
are parameters 
that depend on the expansion coefficient $\psi_{n,\ell}$
as follows: 
\beq
\begin{aligned}
\langle|\ell|\rangle&=
\sum_{\ell=-\infty}^{+\infty}
\sum^\infty_{n=0}
|\ell|\,|\psi_{n,\ell}|^2\,,
\\
\Phi&\equiv\sum_{\ell=-\infty}^{+\infty}
\sum^\infty_{n=0}
2n|\psi_{n,\ell}|^2\,,
\\
\beta
&=\sum_{\ell=-\infty}^{+\infty}
\sum^\infty_{n=0}2\sqrt{n(|\ell|+n)}\psi_{n,\ell}\psi^*_{n-1,\ell}\,.
\end{aligned}
\eeq
Since $w(z)=w_0\sqrt{1+(z/z_0)^2}$ and $\exp[i\zeta(z)]=(z_0+i z)/|z_0+i z|$, by defining
$\alpha=1+\langle|\ell|\rangle+\Phi$, 
it is possible to show that
the variance $\sigma^2_r(z)$ has a parabolic
dependence on $z$, namely:
\beq
\begin{aligned}
\sigma^2(z)
=&\sigma^2_m+\theta^2_{\rm rms}(z-z_m)^2\,,
\end{aligned}
\eeq
with
\beq
\begin{aligned}
z_m&=-z_0\frac{\Im{\rm m}\beta}{\Re{\rm e}(\alpha+\beta)}\,,
\\
\sigma^2_m&=\frac{w^2_0}{2}\frac{\alpha^2-|\beta|^2}{\Re{\rm e}(\alpha+\beta)}\,,
\\ 
\theta_{\rm rms}&=\frac{w_0}{\sqrt2z_0}\sqrt{\Re{\rm e}(\alpha+\beta)}\,.
\end{aligned}
\eeq

Since the transverse scale can be fixed by $w_0$ or
equivalently by $\sigma_m$, we can express 
$\theta_{\rm rms}$ and $z_m$ in function of $\sigma_m$, obtaining
	\beq
\begin{aligned}
\theta_{\rm rms}&=
\frac{1}{k\sigma_m}\sqrt{\alpha^2-|\beta|^2}\,,
\\
z_m&=-k \sigma^2_m \frac{\Im{\rm m}\beta}{\alpha^2-|\beta|^2}
=-\frac{\Im{\rm m}\beta}{k\theta_{\rm rms}^2}\,.
\end{aligned}
\eeq

We have thus demonstrated eqs. (5) and (6)
of the main text.

\section{Proof of the bound}
\label{appendixB}
In this section we demonstrate that the following bound holds:
\beq
\label{bound_SI}
\begin{aligned}
\mathcal M_{\rm rms}^2\equiv k\theta_{\rm rms}\sigma_m\geq1+\langle|\ell|\rangle\,.
\end{aligned}
\eeq
In the previous section we have
shown that, for a generic beam, $k\theta_{\rm rms}\sigma_m=\sqrt{\alpha^2-|\beta|^2}$,
with
$\alpha=\sum_\ell\sum_n(1+2n+|\ell|)|\psi_{n,\ell}|^2$
and
$\beta=\sum_\ell
\sum_n2\sqrt{n(|\ell|+n)}\psi_{n,\ell}\psi^*_{n-1,\ell}$.
Our goal is 
to find a lower bound for $\sqrt{\alpha^2-|\beta|^2}$ depending on the average value of the OAM.

If we define $\widetilde\beta=\sum_{\ell}\sum_{n}2\sqrt{n(|\ell|+n)}|\psi_{n,\ell}\psi_{n-1,\ell}|$,
by the properties of the absolute value, it follows that
$|\beta|\leq\widetilde\beta$ and:
\beq
\label{first_bound_SI}
\alpha^2-|\beta|^2\geq\alpha^2-\widetilde\beta^2\,.
\eeq

We now define 
$\Delta^{\pm}=\sum_\ell\sum^\infty_{n=0}(\sqrt{|\ell|+n+1}|\psi_{n,\ell}|\pm\sqrt{n+1}|\psi_{n+1,,\ell}|)^2$.
By expanding the square in $\Delta^\pm$ it possible to show that
\beq
\begin{aligned}
\Delta^\pm=\alpha\pm\widetilde \beta\,.
\end{aligned}
\eeq
Then
\beq
\label{D+D-_SI}
\alpha^2-\widetilde\beta^2=
(\alpha+\widetilde\beta)(\alpha-\widetilde\beta)=\Delta^+\Delta^-\,.
\eeq

\begin{figure}[t]
\includegraphics[width=8cm]{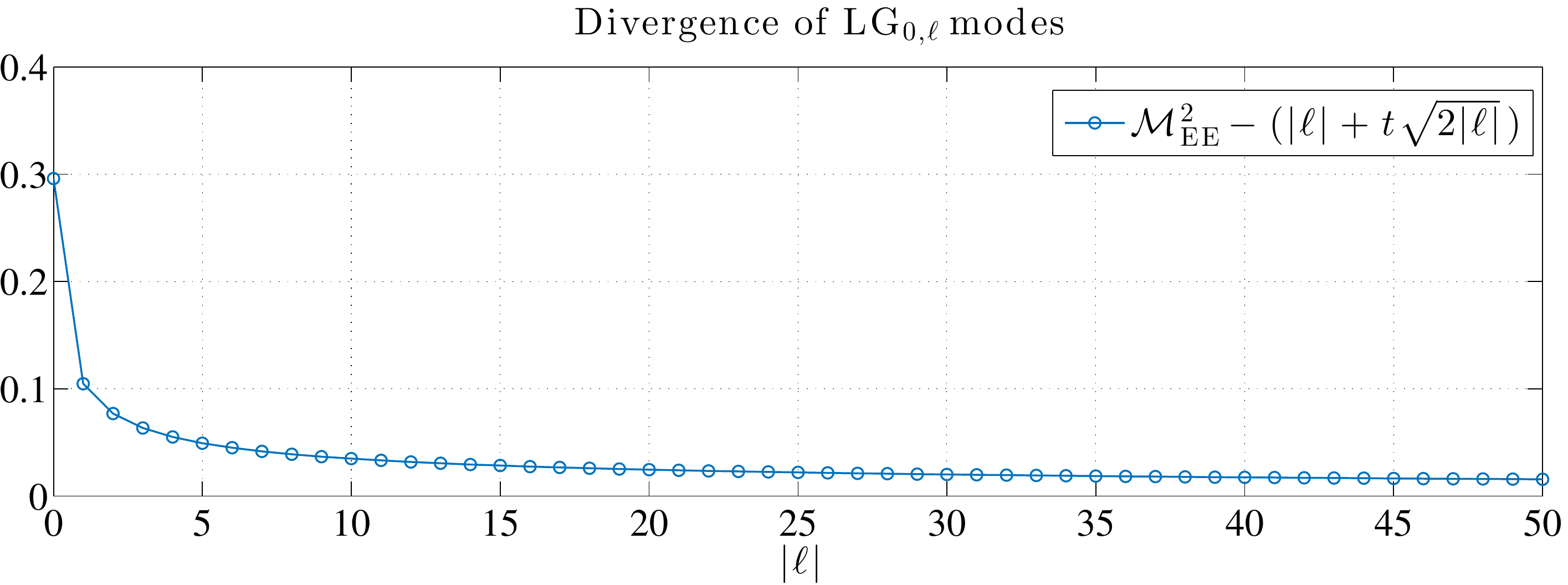}
\caption{(Color online) Difference between the $\mathcal M^2_{\rm EE}$
value and its approximaton, $|\ell|+t\sqrt{2|\ell|}$, for
the LG$_{0,\ell}$ modes.
}
\label{fig_divLG0}
\end{figure}

We note that $\Delta^\pm$ can be written as the expectation value of operators in a real valued
vector space. Let's consider an infinite
dimensional space spanned by the orthonormal vectors
$\{\ket{n,\ell}\}$ with $n=0,\cdots,+\infty$ and
$\ell=-\infty,\cdots,+\infty$ and define the following 
operators $\hat N$, $\hat L$, $\hat a$ and $\hat a^\dag$
by their
action on the basis states: 
\begin{align}
\hat N\ket{n,\ell}&=n \ket{n,\ell},
&\qquad&
\hat a\ket{n,\ell}=\sqrt{n}\ket{n-1,\ell},
\\
\notag\hat L\ket{n,\ell}&=\ell \ket{n,\ell},
&&
\hat a^\dag\ket{n,\ell}=\sqrt{n+1}\ket{n+1,\ell}.
\end{align}
If we define 
$\hat A_\pm=\sqrt{\hat N+|\hat L|+1}\pm \hat a$ 
and a generic vector $\ket v$ as $\ket v=\sum_l\sum_n |\psi_{n,\ell}|\ket {n,\ell}$, we have
\beq
\begin{aligned}
\hat  A{_\pm}\ket{v}
&=
\sum_{n,\ell}
\left[\sqrt{n+|\ell|+1}|\psi_{n,\ell}|\pm\sqrt{n+1}
|\psi_{n+1,\ell}|\right]\ket{n,\ell}
\end{aligned}
\eeq
Since the vectors $\ket{n,\ell}$ are orthonormal
and thus satisfy 
$\braket{n',\ell'}{n,\ell}=\delta_{n,n'}\delta_{\ell,\ell'}$
the expectation value of 
$A^\dag_\pm A_\pm$ are precisely $\Delta^\pm$:
\beq
\begin{aligned}
\bra v A^\dag_\pm A_\pm\ket v=
\Delta^\pm\,.
\end{aligned}
\eeq
By the triangular inequality we may then bound 
$\Delta^+\Delta ^-$:
\beq
\label{boundD+D-_SI}
\Delta^+\Delta^-=\bra v A^\dag_+ A_+\ket v
\bra v A^\dag_- A_-\ket v
\geq 
|\bra v A^\dag_+A_-\ket v|^2
\eeq
The r.h.s of the previous equation can be explicitely 
evaluated to give:
\beq 
\label{A+A-}
|\bra v A^\dag_+A_-\ket v|^2=
(1+\langle|\ell|\rangle)^2\,.
\eeq 
By combining eqs. \eqref{first_bound_SI}, \eqref{D+D-_SI},
\eqref{boundD+D-_SI} and \eqref{A+A-},
it follows that 
\beq
\begin{aligned}
k\theta_{\rm rms}\sigma_m\geq
1+\langle|\ell|\rangle
\end{aligned}
\eeq

\begin{figure}[tbp!]
\includegraphics[width=8cm]{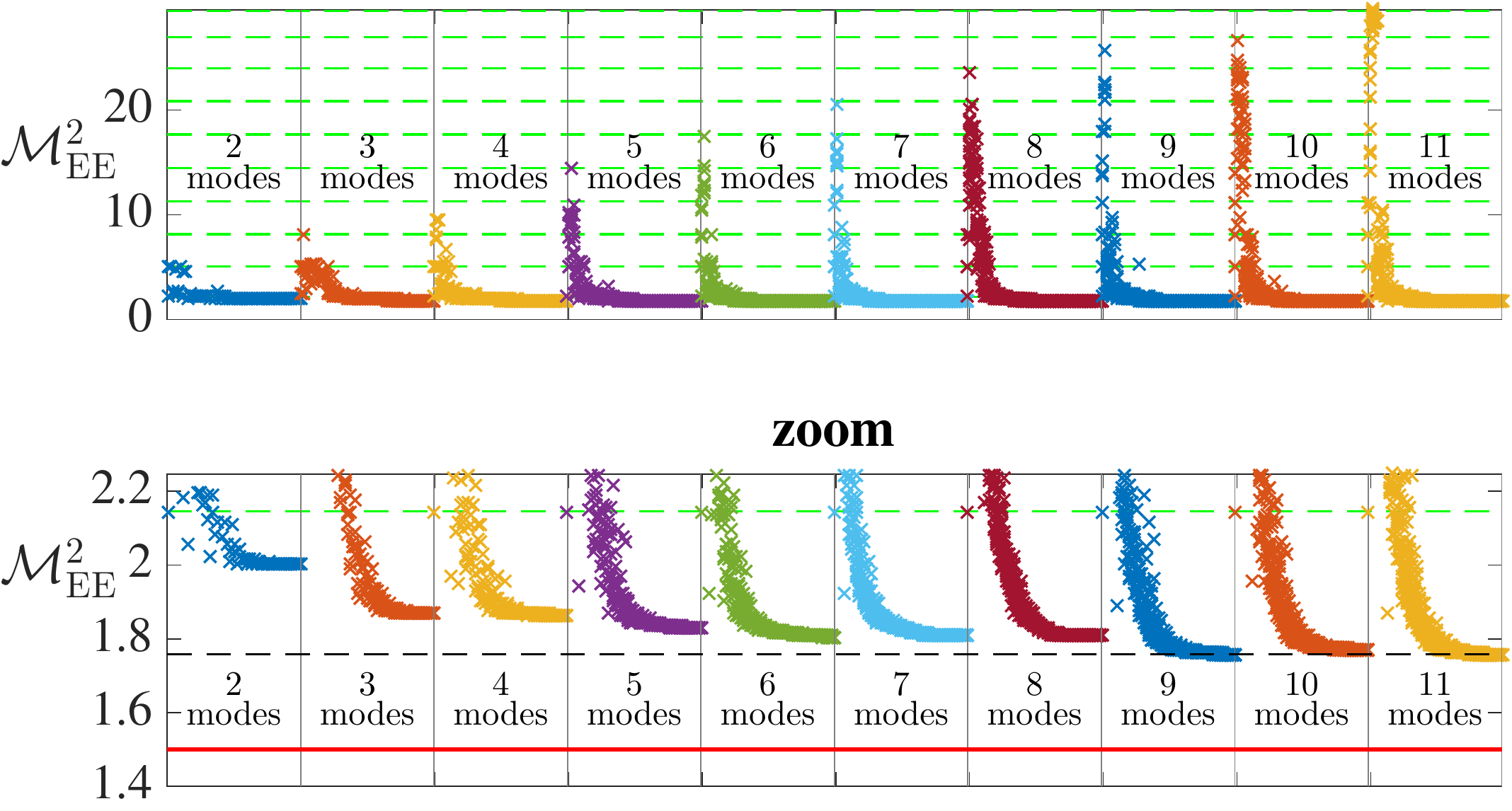}
\caption{(Color online) Steps of the Nelder-Mead used to minimize $\MEE$. We used $\ell=1$ and $E_0\simeq0.63$. We report the
number of LG modes used in the superposition
of eq. \eqref{truncated_sup}. With green (light gray) dashed
lines we show the $\MEE$ values of the
LG$_{n,1}$ modes, while the black dashed line
is the minimum $\MEE$ achieved.
The continuous red (dark gray) line represents the bound $c_0+|\ell|$ with $c_0=0.5$.
}
\label{minimization}
\end{figure}

\section{Encircled-energy divergence}
\label{appendixC}
Here we evaluate the 
encircled-energy divergence in function
of the expansion coefficient $\psi_{n,\ell}$.
We consider a generic paraxial beam 
expanded as a superposition of LG modes, 
as in equation \eqref{LGexp_SI}.
The integral 
$\int\dd\phi \int^{R_{\rm EE}(z)}_0\dd r\, r\, I(x,y,x)=E_0$
defining the encircled-energy
radius $R(z)$ can be written 
in adimensional notation as
\beq
\label{Tgeneral}
E_0=\int_0^{T(Z)}\sum_\ell
|\mathcal U_\ell(t,Z)|^2\dd t\,,
\eeq
with $Z=z/z_0$ and
\beq
\label{TEE}
\begin{aligned}
\mathcal U_\ell(t,Z)&=
\sum_{n=0}^\infty
\sqrt{\frac{e^{-t}t^{|\ell|} n!}{(|\ell|+n)!}} \psi_{n,\ell}
\left(\frac{1+iZ}{1-iZ}\right)^{n}
L_n^{(|\ell|)}(t)
\\
T(Z)&=\frac{2R^2_{\rm EE}(Zz_0)}{w_0^2(1+Z^2)}\,.
\end{aligned}
\eeq

The minimum $R_m$ of $R_{EE}(z)$ should be found by minimizing the function 
$\sqrt{(1+Z^2)T_{\rm EE}(Z)}$.
We define $T_\infty\equiv\lim_{Z\rightarrow \infty}T(Z)$, namely
\beq
\label{Tinf}
E_0=\int_0^{T_\infty}\sum_\ell|\mathcal U^\infty_\ell(t)|^2\dd t\,,
\eeq
 and 
$\mathcal U^\infty_\ell(t)=
\sqrt{e^{-t}t^{|m|}}\sum_{n=0}^N \sqrt{\frac{ n!}{(|m|+n)!}} \psi_n (-1)^{n}L_n^{(m)}(t)
$. By using the definition of $T(Z)$ 
in eq. \eqref{TEE} it is possible to show that
$T_\infty=\frac{2z_0^2}{w_0^2}\thEE^2$.
By using $z_0=kw_0^2/2$ and $R_m/w_0=
\sqrt{\min_Z[(1+Z^2)T(Z)]/2}$ we may express
$\thEE$ as 
\beq
\begin{aligned}
\thEE&=\frac{\sqrt2}{kw_0}\sqrt{T_\infty}
\\
&=\frac{1}{k R_m}\sqrt{{T_\infty}{\min_Z[(1+Z^2)T(Z)]}}\,.
\end{aligned}
\eeq
Also in this case
the product $\mathcal M^2_{\rm EE}\equiv k\thEE R_m$ is adimensional and
depends only on the expansion coefficient $\psi_{n,\ell}$.

The above relations can be simplified for LG modes.
Indeed, for an LG mode, $T(Z)$ is an
even function of $Z$ and its minimum
is obtained at $Z=0$. Moreover we have 
$T(Z)=T(0)=T_\infty\equiv T_{n,\ell}$.
Then the divergence
for a LG$_{n,\ell}$ mode can be expressed as
\beq
{\thEE}_{n,\ell}=\frac{1}{k R_m}T_{n,\ell}\,,
\eeq
with
\beq
\label{TLG}
E_0=\frac{n!}{(|\ell|+n)!}\int_0^{T_{n,\ell}}e^{-t}t^{|\ell|}(L^{(|\ell|)}_n(t))^2\dd t\,.
\eeq

\begin{figure}[tbp!]
\includegraphics[width=8cm]{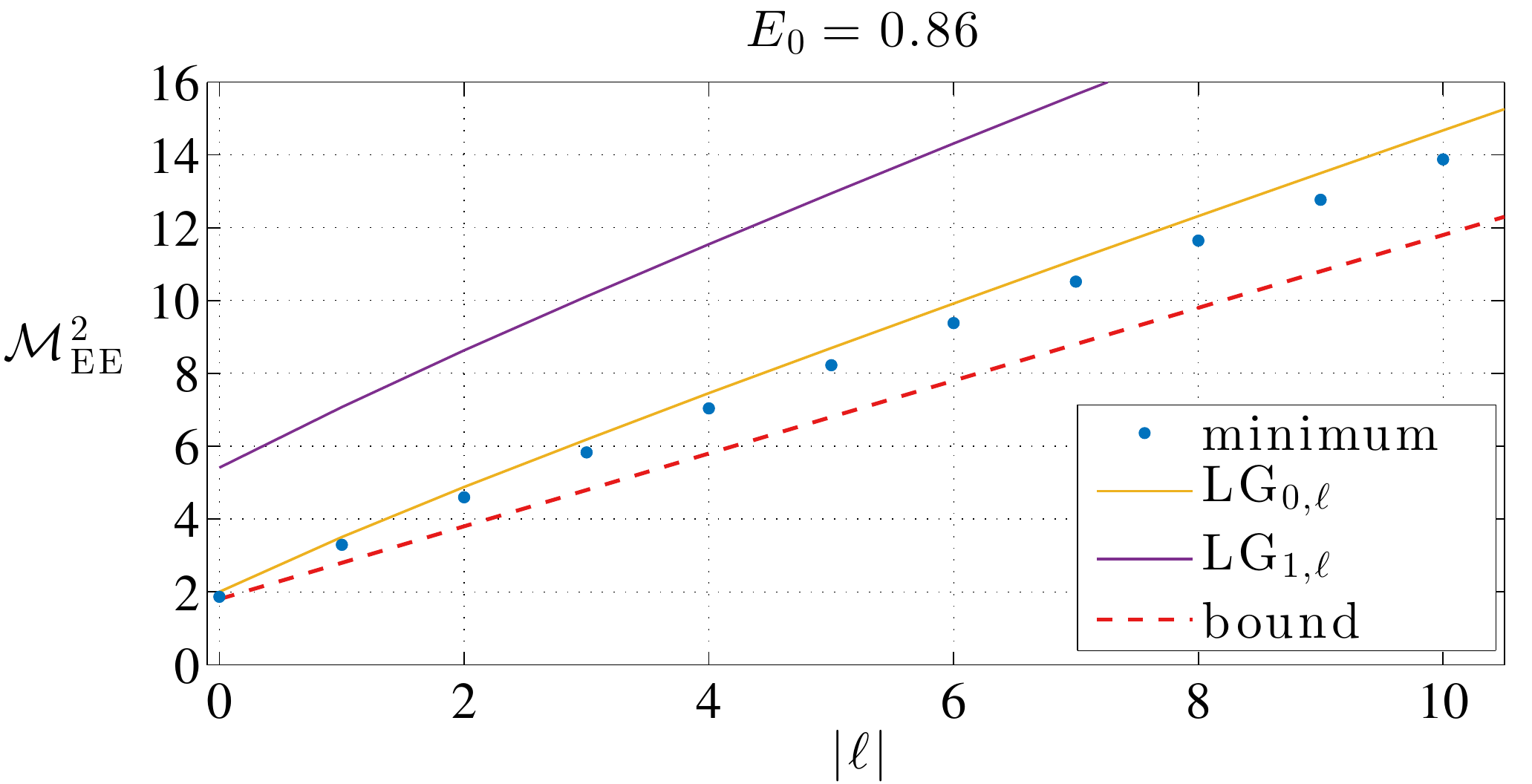}
\includegraphics[width=8cm]{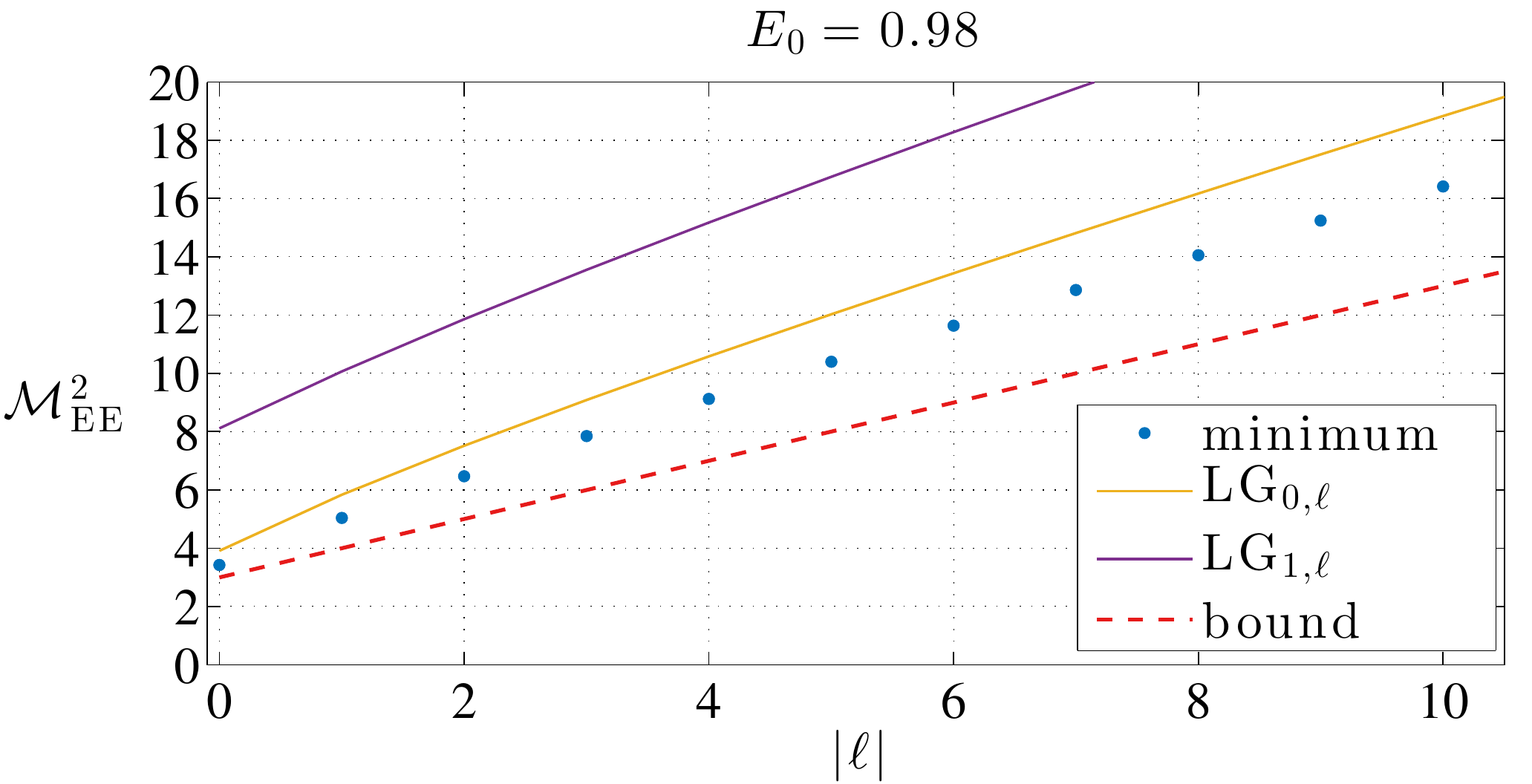}
\caption{(Color online)  Minimum values of $\MEE$ calculated
over beams with fixed value of $\ell$.
We used $E_0\simeq 0.86$ in the
upper graph and $E_0=0.98$ in the
lower graph. 
}
\label{kthR_SI}
\end{figure}
The divergence of the LG$_{0,\ell}$ modes can be further
simplified and it is expressed
through special functions: in this case, equation \eqref{TLG} can
written as $P(|\ell|+1,T_{0,\ell})=E_0$, where
$P(a,z)$ is the regularized incomplee Gamma function
$P(a,z)=\frac{1}{\Gamma(a)}\int^z_0t^{a-1}e^{-t}\dd t$.
By using a formula due to Tricomi \cite{tric50mze} the
asymptotic behavior at large $|\ell|$ of $P(a,z)$ is the following:
\beq
P(|\ell|+1,|\ell|+t\sqrt{2|\ell|} )=\frac12[|1|+{\rm erf}(t)]+O(\frac{1}{\sqrt{|\ell|}})\,,
\eeq
with ${\rm erf}(t)$ the error function.
Then, at large $|\ell|$, the asymptotic divergence of the
LG$_{0,\ell}$ mode is
\beq
\label{approx_SI}
\theta^{(EE)}_{0,\ell}\sim \frac{1}{kR_m}(|\ell|+ t\sqrt{2|\ell|})\,,
\eeq
with $t={\rm erf}^{-1}(2E_0-1)$.
Note that when $E_0=\frac12 +\epsilon$ we have $t\sim\sqrt{\pi}\epsilon$.
The r.h.s. of eq. \eqref{approx_SI} is a good
approximation of the divergence also for low values of $|\ell|$, as demonstrated
by Fig. \ref{fig_divLG0}.

\begin{figure}[tbp!]
\includegraphics[width=7.8cm]{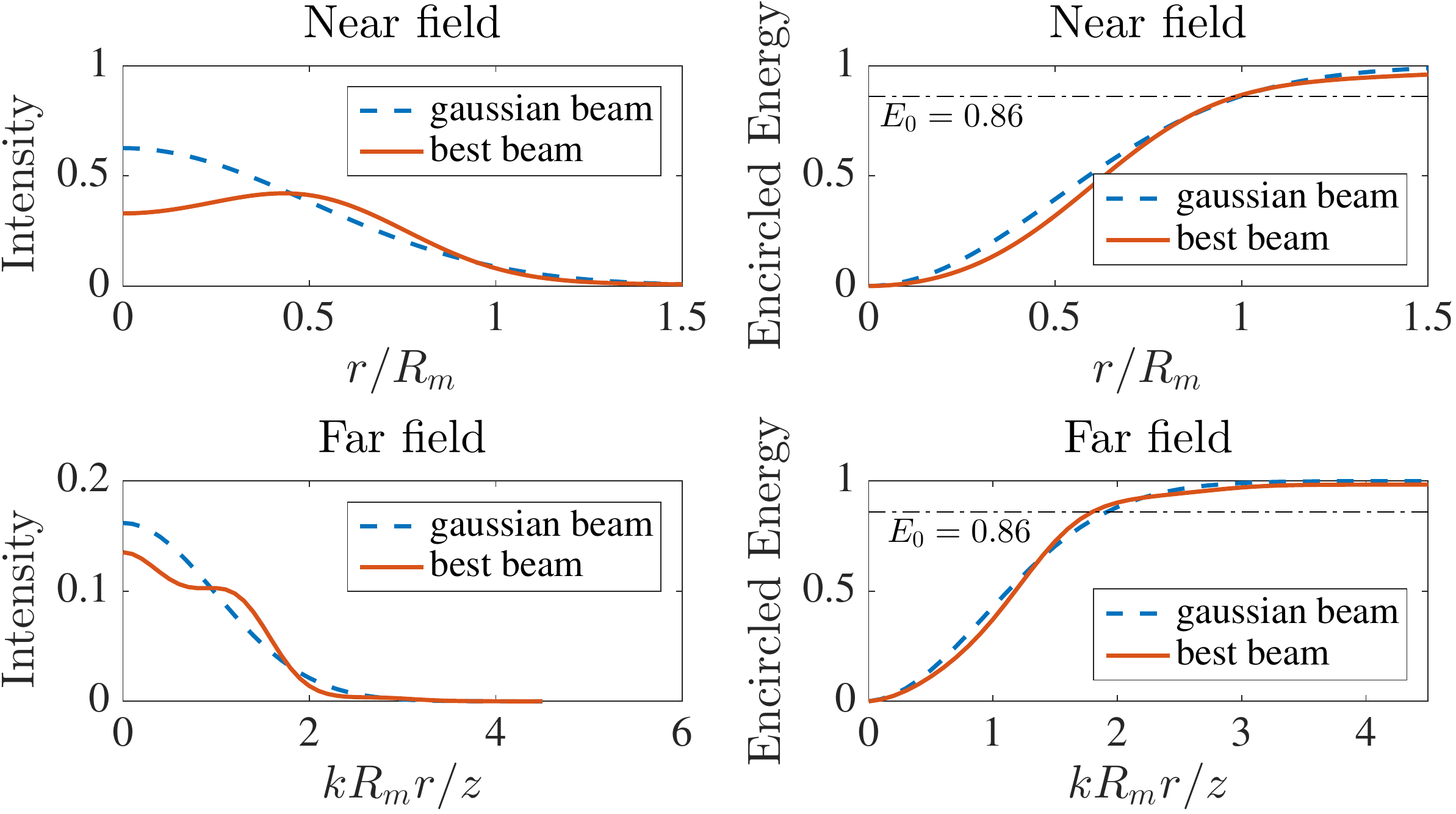}
\hskip1cm 
\includegraphics[width=7.8cm]{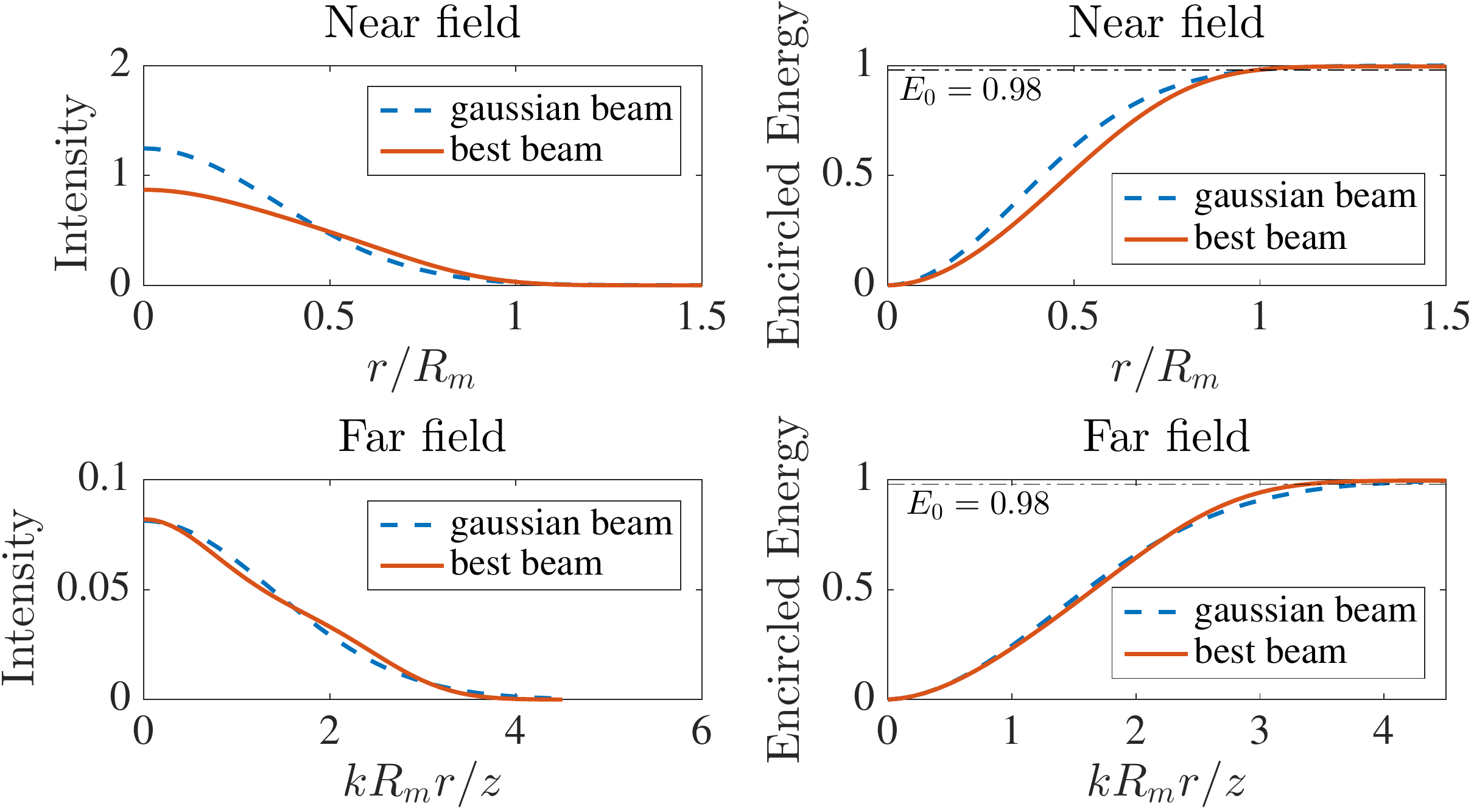}
\caption{(Color online)  Intensity patterns and encircled-energy for the Gaussian beam and the best beam at $\ell=0$ that
minimize $\mathcal M^2_{\rm EE}$. The physical size of the beams are chosen such that they have the same value of $R_m$.
With dash-dot lines we indicate the value of $E_0$ that define the encircled-energy radius $R(z)$. 
The optimal beam is obtained for $E_0=0.86$ for the two upper rows figures, while  $E_0=0.98$ in the two bottom  rows.
}
\label{bestEE_SI}
\end{figure}

\section{Numerical minimization}
\label{appendixD}
As described in the main text, we numerically evaluated the minimum of 
$\MEE$ at fixed value of OAM $\ell$. Here we illustrated with more details
the method.
For each value of $\ell$,
we defined the generic wavefunction as a truncated superposition of the first $N$ 
${\rm LG}_{n,\ell}$ modes:
\beq 
\label{truncated_sup}
\Psi_\ell(\{\psi_n\})=\sum_{n=0}^{N-1}\psi_n{\rm LG}_{n,\ell}(r,\phi,z)\,.
\eeq 
The wavefunction is uniquely determined by
the coefficients $\{\psi_0,\psi_1,\cdots,\psi_{N-1}\}$. Due to the equivalence under a global phase,
we set $\psi_0\in \mathbb R$, while we considered the remaining coefficients as complex. 
By using the Nelder-Mead algorithm~\cite{neld65tcj},
we search for the coefficients $\{\psi_n\}$ that
minimize the $\MEE$. To take into account
the normalization $\sum_n|\psi_n|^2=1$,
we adapted the algorithm to an hypersphere.
We here recall that the $\MEE$
parameter is defined in eq. (13) of the main text.
In fig. \ref{minimization}  we show the iterations of the algorithm for
different truncation $N$
and for $\ell=1$ and $E_0\simeq0.63$.
By increasing the
number of modes, the minimum value of $\MEE$ decreases
but it is always larger than the bound $c_0+|\ell|$
with $c_0=0.5$.
The procedure was repeated for different values
of $\ell$ to obtain the graph shown in fig. 3 in the main text. 

We also performed the minimization
for different values of $E_0$. The
results are presented in
Fig. \ref{kthR_SI} for $E_0\simeq0.86$ and $E_0=0.98$.
In both cases the bound $c_0+|\ell|$ holds with
$c_0=1.8$ and $c_0=3$ respectively.

{ Finally, in Fig. \ref{bestEE_SI}, for $\ell=0$ we show the intensity patterns of the best divergence beam and the
comparison with the gaussian beam (namely the lowest LG mode)  and $E_0=0.86$ and $E_0=0.98$.}


%
%
%
%

\begin{thebibliography}{10}
\bibitem{foot2}
We use a different notation
with respect to \cite{phil83apo},
where the beam radius is defined
as $\sqrt{2}\sigma_r(z)$.
\end{thebibliography}
%

%

\end{document}